\documentclass[fleqn,10pt]{wlscirep}
\usepackage[T1]{fontenc}
\usepackage[separate-uncertainty=true]{siunitx}
\usepackage{glossaries}
\glsdisablehyper
\usepackage{algorithm}
\usepackage{algpseudocode}

\newacronym{FDA}{US~FDA}{United States Food and Drug Administration}
\newacronym{LD}{LD}{Laser Diode}
\newacronym{LED}{LED}{Light Emitting Diode}
\newacronym{PPG}{PPG}{PhotoPlethysmoGram}
\newacronym{ABG}{ABG}{Arterial Blood Gas}
\newacronym{mBLL}{mBLL}{modified Beer-Lambert Law}
\newacronym{SaO2}{\ensuremath{\text{SaO\textsubscript{2}}}}{arterial Oxygen Saturation}
\newacronym{SpO2}{\ensuremath{\text{SpO\textsubscript{2}}}}{pulsatile Oxygen Saturation}
\newacronym{StO2}{\ensuremath{\text{StO\textsubscript{2}}}}{tissue Oxygen Saturation}
\newacronym{HbO2}{\ensuremath{\text{HbO\textsubscript{2}}}}{OxyHemoglobin}
\newacronym{Hb}{\ensuremath{\text{Hb}}}{deoxyHemoglobin}
\newacronym{HbT}{\ensuremath{\text{HbT}}}{Total Hemoglobin}
\newacronym{IR}{IR}{InfraRed}
\newacronym{RoR}{\ensuremath{\text{RoR}}}{Ratio-of-Ratios}
\newacronym{ITA}{ITA}{Individual Typology Angle}
\newacronym{M}{\ensuremath{M}}{epidermal Melanin volume fraction}
\newacronym{W}{\ensuremath{W}}{Water volume fraction}
\newacronym{HR}{\ensuremath{f_{HR}}}{heart rate}
\newacronym{FWHM}{FWHM}{Full-Width at Half-Maximum}
\newacronym{DOF}{DoF}{Degrees of Freedom}
\newacronym{mua}{\ensuremath{\mu_a}}{absorption coefficient}
\newacronym{musp}{\ensuremath{\mu_s^\prime}}{reduced scattering coefficient}
\newacronym{rho}{\ensuremath{\rho}}{source--detector separation}

\title{Melanin- and linewidth-corrected pulse-oximetry}

\author[1,*]{Giles~Blaney}
\author[1]{Jodee~Frias}
\author[2]{Ravi~Durbha}
\author[1]{Sergio~Fantini}
\author[2]{Valencia~Koomson}
\affil[1]{Biomedical Engineering, Tufts University, Medford, MA, USA}
\affil[2]{Electrical and Computer Engineering, Tufts University, Medford, MA, USA}
\affil[*]{Giles.Blaney@tufts.edu}

\keywords{Pulse oximetry, skin tone bias, melanin, spectral coloring, modified Beer-Lambert law, OpenOximetry}

\begin{abstract} 
	Current pulse oximeters tend to overestimate arterial saturation in darkly pigmented patients, increasing occult hypoxemia rates. 
	One proposed mechanism is spectral coloring: melanin contributes to reshaping the detected spectrum of a broad-linewidth source, making the wavelength-dependent parameters behind calibration skin-tone-dependent. 
	Diffuse optics theory allows for a melanin- and linewidth-corrected calibration equation, which to our knowledge has never been tested in vivo. 
	In this work, we fit the empirical linear equation, the diffuse optics theoretical equation, and its corrected form to the public OpenOximetry Repository---considering 98 patients with paired arterial blood-draws, raw photoplethysmograms, and skin reflectance spectra. 
	Accuracy is practically indistinguishable between the three approaches, yet the saturation error’s dependence on melanin falls from +2.8 to +1.2 to +0.1 percentage points per unit melanin volume fraction---consistent with the spectral-coloring theory. 
	None of these dependencies on melanin are statistically resolved here---a cohort roughly ten times larger is needed---so the correction is backed by theory and its expected direction. 
	All three calibration approaches carry two degrees of freedom: the theoretical equation drops into existing calibration procedures, and its melanin-aware form needs a per-patient skin-tone measurement. 
	This work demonstrates how the theoretical pulse oximetry model can be applied in calibration. 
\end{abstract}

\begin{document}

\flushbottom
\maketitle
\thispagestyle{empty}

\section*{Introduction}
Pulse oximetry is a non-invasive optical technique that aims to measure a patient's \gls{SaO2} using red and \gls{IR} light that has traversed tissue.
The index recovered by a pulse oximeter is typically called the \gls{SpO2} (n.b., the ``p'' in \gls{SpO2} can denote ``peripheral'' as well as ``pulsatile''); however, in this work we will use \gls{SaO2} since it represents the target quantity.
Pulse oximeters focus on measuring the ratio of relative pulsatile amplitudes---from cardiac pulsation---at red versus \gls{IR}, called the \gls{RoR}\cite{Aoyagi_JAnesth03_PulseOximetry,Charlton_Proc.IEEE22_WearablePhotoplethysmography}.
Connecting the measured \gls{RoR} to \gls{SaO2} is typically achieved using an empirical calibration study on a relatively small cohort, on the order of \qty{10}{patients}\cite{FDACDRH_13_PulseOximeters}.
However, in recent years it has become clear that current pulse oximetry devices exhibit a skin-tone-correlated bias causing erroneously high \gls{SaO2} readings for Black versus White patients\cite{SjodingMichaelW._NEJM20_RacialBias,Shi_BMCMed.22_AccuracyPulse,Bickler_Ane.22_PulseOximeter,Cabanas_Sensors22_SkinPigmentation,Al-Halawani_Physiol.Meas.23_ReviewEffect,Martin_BJA24_EffectSkin}.
A consequence of this bias is higher rates of occult hypoxemia in patients with darker skin pigmentation\cite{Chesley_RC22_RacialDisparities,Gudelunas_AA24_LowPerfusion,Hendrickson_CHESTCriticalCare26_EquiOxProspective} (n.b., \emph{occult hypoxemia} occurs when true \gls{SaO2} is low enough to warrant intervention, but intervention is not offered since the oximeter incorrectly indicates a high \gls{SaO2}).
The current \gls{FDA} calibration protocol from 2013 recommends at least two darkly pigmented subjects, or \qty{15}{\percent} of the calibration cohort, whichever is larger\cite{FDACDRH_13_PulseOximeters}.
This protocol itself is under active reconsideration, with a 2024 \gls{FDA} advisory panel reviewing the 2013 clinical study design and weighing pigmentation scales---including the \gls{ITA} considered here---for performance evaluation\cite{FDAARTDP_24_FDAExecutive}.
This asymmetric cohort requirement may explain the skin-tone bias, but raises the question of why the calibration would depend on skin tone and how calibration can be modified to account for it.
\par

One proposed mechanism for the skin-tone bias is spectral coloring of broad-linewidth light sources\cite{Rea_BJA23_LightSource,Bierman_BJA24_MelaninBias,Benner_JBO26_CauseEffect,Benner_BritishJournalofAnaesthesia25_LightSource,Blaney_JBO26_BroadlinewidthSources}.
The argument is that if broad-linewidth sources are used (e.g., \glspl{LED}) as opposed to monochromatic sources (e.g., \glspl{LD}), the detected light spectrum and its nominal wavelength depend on the tissue optical properties, including a strong dependence on \gls{M}.
Ultimately, this means that the physical constants that depend on wavelength (e.g., the hemoglobin extinction coefficients) need to be adjusted for different skin tones, necessitating \gls{M}-dependent calibration equations.
This mechanism has recently received direct experimental support: substituting narrow-linewidth \glspl{LD} for the \glspl{LED} of a conventional oximeter removed the pigmentation-related bias in both a benchtop test and a clinical study of \qty{18}{participants}, while the two \gls{LED}-based devices tested exhibited a bias\cite{Pologe_PLOSONE25_LaserbasedPulse}.
Alternative mechanisms causing the skin-tone bias have also been proposed, such as a changing dermal sensitivity with skin tone\cite{Blaney_JBO24_DualratioApproach} or non-linear effects involving changes in the background optical properties during the cardiac cycle interacting with \gls{M}\cite{Al-Halawani_JBO24_MonteCarlo}.
It is unclear whether one particular mechanism is the dominant source of the skin-tone bias, and it is possible that the bias is in fact the result of a combination of these mechanisms.
\par

The empirical nature of pulse oximeter calibration itself has also been called into question, since the calibration is somewhat blind to what may confound it and how its recovered coefficients relate to physical quantities\cite{Stuban_PPEE08_NoninvasiveCalibration,Chan_RespiratoryMedicine13_PulseOximetry}.
Recently, we derived the linear pulse oximeter calibration equation from the \gls{mBLL}, allowing one to relate the linear coefficients to physical quantities\cite{Blaney_JBO24_CriticalAnalysis}.
This work also allowed us to arrive at a non-linearized theoretical expression for the relationship between \gls{SaO2} and \gls{RoR} based on hemoglobin molar extinction coefficients and optical path-lengths.
We further tested the spectral coloring mechanism for skin-tone bias on the basis of this derivation, entirely in simulation, and initially proposed a spectral-coloring correction to pulse oximeter calibration\cite{Blaney_JBO26_BroadlinewidthSources}.
This correction may now be tested on in vivo pulse oximeter calibration data.
Such a test requires raw red and \gls{IR} \gls{PPG} traces paired with \gls{ABG} \gls{SaO2} in a skin-tone-diverse population, which the public OpenOximetry Repository provides\cite{Fong_24_OpenOximetryRepository}.
An important aspect of the proposed correction is a quantitative measure of skin tone directly related to \gls{M}, which may also be derived from the OpenOximetry Repository's raw skin reflectance spectra.
\par

The core of this work considers three calibration cases: first, the standard linear empirical calibration; second, the \gls{mBLL}-derived expression with an extra additive term; and third, the same \gls{mBLL}-based expression including a correction for spectral coloring based on \gls{M}.
Importantly, in our investigation we consider only calibration methods with two \gls{DOF}, so that the three cases are compared fairly---for this reason we do not include the sometimes-used empirical quadratic calibration, which would require three \gls{DOF}.
To our knowledge, this work therefore constitutes the first investigation of the \gls{mBLL}-derived pulse oximeter calibration equation\cite{Blaney_JBO24_CriticalAnalysis} and its spectral-coloring, skin-tone-dependent correction\cite{Blaney_JBO26_BroadlinewidthSources} on an experimental controlled-desaturation dataset where \gls{M} may also be quantitatively determined (i.e., the OpenOximetry Repository\cite{Fong_24_OpenOximetryRepository}).
\par

\section*{Results}
\subsection*{OpenOximetry analyzable cohort}
\begin{figure}[!htb]
	\glsreset{ABG}\glsreset{SaO2}\glsreset{PPG}\glsreset{RoR}\glsreset{M}
	\begin{center}
		\includegraphics{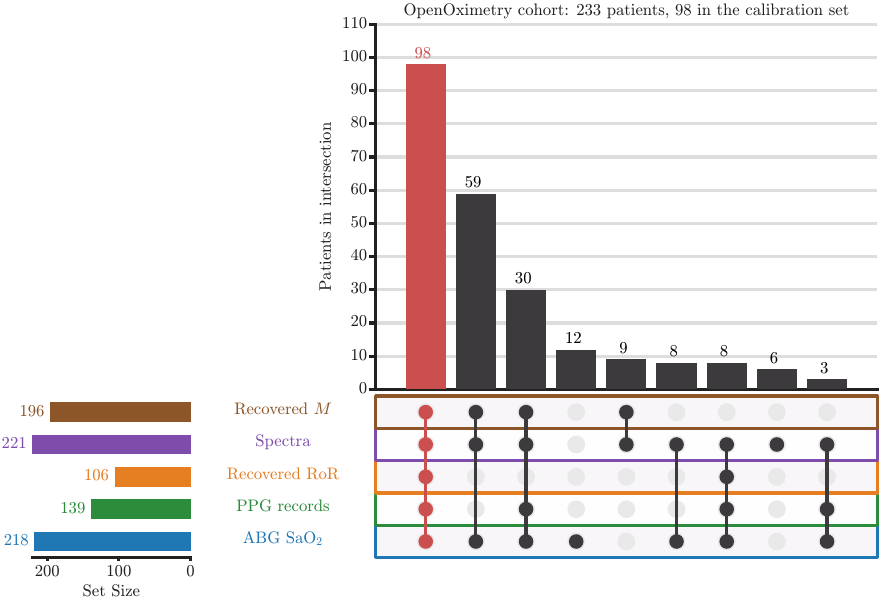}
	\end{center}
	\caption{
		UpSet plot of the OpenOximetry cohort divided into sets relevant for this work. 
		All values refer to numbers of patients and the sum of all intersection bars represents the total number in the dataset (\qty{233}{patients}).
		(Blue)~\qty{218}{patients} have \gls{SaO2} values from \gls{ABG}.
		(Green)~\qty{139}{patients} have raw \gls{PPG} records.
		(Orange)~\Gls{RoR} values were recovered for \qty{106}{patients}.
		(Purple)~\qty{221}{patients} have reflectance spectra.
		(Brown)~Finger \gls{M} values were recovered for \qty{196}{patients}.
		(Red)~\qty{98}{patients} satisfy the intersection of the three sets needed for calibration (\gls{ABG} \gls{SaO2}, Recovered \gls{RoR}, and Recovered \gls{M}).
	}
	\label{fig:cohort}
\end{figure}

\subsubsection*{Oximetry parameters}
The two oximetry parameters required for calibration are \gls{SaO2} and \gls{RoR}. 
Of the total \qty{233}{patients} in the dataset, \qty{106}{patients} satisfied this requirement (Figure~\ref{fig:cohort} -- orange set)---of which no \gls{M} was recovered for \qty{8}{patients}, leaving \qty{98}{patients} in this work's calibration set (Figure~\ref{fig:cohort} -- red intersection).
The \gls{ABG} \gls{SaO2} requirement is the least restrictive since almost every patient in the dataset included these data with \qty{218}{patients} contributing (Figure~\ref{fig:cohort} -- blue set).
The majority of the \qty{135}{patients} excluded from calibration were excluded due to no \gls{PPG} record because the OpenOximetry dataset collected these data on only \qty{139}{patients} (Figure~\ref{fig:cohort} -- green set).
Therefore, the signal processing methods herein excluded $33/139$ or \qty{24}{\percent} of the \gls{PPG} records when recovering \gls{RoR}.
\par

The statistics of the \gls{SaO2} and \gls{RoR} data in the calibration set are reported in Table~\ref{tab:calstats}.
The calibration set's \gls{SaO2} range is narrower than the \qtyrange{58.1}{100.0}{\percent} range of accepted \gls{SaO2} values (n.b., the full dataset includes values \qty{>100}{\percent}, which are not accepted).
This is because very low values of \gls{SaO2} were often associated with no recovered \gls{RoR}. 
Despite this, the calibration set's range extends below the \gls{FDA} lower calibration bound of \qty{70}{\percent}\cite{FDACDRH_13_PulseOximeters}.
\par

\begin{table}[!htb]
	\glsreset{SaO2}\glsreset{RoR}
	\centering
	\small
	\caption{Summary of the calibration set's oximetry parameters (\qty{98}{patients} and \qty{2991}{samples})}
	\label{tab:calstats}
	\begin{tabular}{lr}
		\toprule
		
		\Gls{SaO2} range                & \qtyrange{65.2}{99.4}{\percent} \\
		\Gls{SaO2} inter-quartile range & \qtyrange{76.4}{93.1}{\percent} \\
		\Gls{SaO2} median               & \qty{85.0}{\percent}  \\
		
		\midrule
		
		\Gls{RoR} range                 & \numrange{0.376}{1.740} \\
		\Gls{RoR} inter-quartile range  & \numrange{0.740}{1.211} \\
		\Gls{RoR} median                & \num{0.986} \\
		
		\bottomrule
	\end{tabular}
\end{table}

\subsubsection*{Skin-tone parameters}
\begin{table}[!htb]
	\glsreset{M}\glsreset{ITA}
	\centering
	\small
	\caption{Summary of finger skin-tone or -color parameters for the \num{98} patients included in and the \num{98} excluded from calibration, of the \num{196} with a recovered \gls{M}}
	\label{tab:pigmentstats}
	\begin{tabular}{lrr}
		\toprule
		& Included in calibration & Excluded from calibration \\
		\midrule
		\midrule
		
		\Gls{M} range                & \numrange{0.022}{0.193} & \numrange{0.027}{0.220} \\
		\Gls{M} inter-quartile range & \numrange{0.042}{0.083} & \numrange{0.038}{0.074} \\
		\Gls{M} median               & \num{0.053}             & \num{0.055} \\
		
		\midrule
		\Gls{ITA} range                & \qtyrange{-27.9}{47.3}{\degree} & \qtyrange{-34.9}{43.6}{\degree} \\
		\Gls{ITA} inter-quartile range &   \qtyrange{6.6}{29.4}{\degree} & \qtyrange{11.5}{32.8}{\degree} \\
		\Gls{ITA} median & \qty{21.7}{\degree} & \qty{23.7}{\degree} \\
		
		\bottomrule
	\end{tabular}
\end{table}

\begin{figure}[!htb]
	\glsreset{M}\glsreset{ITA}
	\begin{center}
		\includegraphics{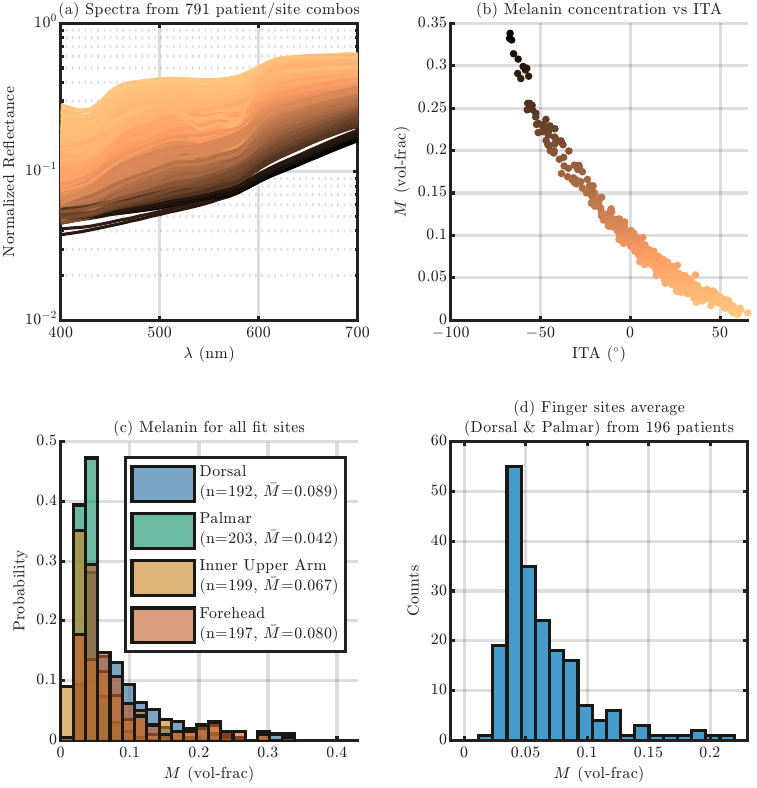}
	\end{center}
	\caption{
		Skin-tone or -color parameters derived from the reflectance spectra.
		(a)~\num{791} spectra representing unique patient-and-site combinations. Line color represents recovered \gls{M}, darker being greater.
		(b)~Recovered \gls{M} versus colorimetric \gls{ITA} for each of the spectra. Point colors match spectra in (a).
		(c)~Distribution of recovered \glspl{M} at each of the four fitted sites.
		(d)~The finger-site pooled \gls{M} values used in further analysis, one value for each of the \num{196} patients with recoverable \gls{M}.
		}
	\label{fig:spectra_ITA_M}
\end{figure}

Skin tone was parameterized by \gls{M} and skin color by \gls{ITA}, both of which were derived from reflectance spectra in the visible range (Figure~\ref{fig:spectra_ITA_M}(a)).
These are not interchangeable---\gls{M} is a spectroscopic concentration recovered through a dermis-optics model, whereas \gls{ITA} is a perceptual summary of that same spectrum.
Both are reported due to the popularity of \gls{ITA}, but only \gls{M} is treated as a physical quantity.
\Gls{M} is plotted versus \gls{ITA} in Figure~\ref{fig:spectra_ITA_M}(b) showing a non-linear relationship.
This relationship between \gls{M} and \gls{ITA} is largely due to \gls{ITA} being an indirect measure of the perceived lightness while \gls{M} is the concentration of a pigment, therefore in the limit of $\text{\acrshort{ITA}}=\qty{-90}{\degree}$ (i.e., completely black), \gls{M} would blow up to infinity (i.e., completely absorbing).
\par

The spectral fit was executed for all \num{791} spectra representing unique patient and site combinations (n.b., this included non-finger sites).
This estimated \num{795} free parameters from the \num{791} spectra in one least-squares minimization (i.e., a free \gls{M} for each spectrum and a free \gls{HbT}, \gls{StO2}, Rayleigh scattering fraction, and specular reflectance factor shared across all spectra), returning shared values of \qty{34}{\micro\mole\per\liter} for \gls{HbT}, \num{0.76} for \gls{StO2}, a Rayleigh scattering fraction of \num{0.52}, and \num{0.087} for the specular reflectance factor, with a coefficient of determination for the entire fit of \num{0.97}.
\Gls{M} recovered values in the range \numrange{0.0068}{0.34} across all spectra (Figure~\ref{fig:spectra_ITA_M}(c)).
However, only values for finger dorsal and palmar sites were continued into the pulse-oximetry analysis, as those were the sites relevant to the pulse oximeter finger clip measurement (Figure~\ref{fig:spectra_ITA_M}(d) and Table~\ref{tab:pigmentstats}).
\par

The 2013 \gls{FDA} guidance imposes a darkly pigmented quota on the calibration set without defining how to determine that criterion\cite{FDACDRH_13_PulseOximeters}.
Therefore, we assumed that this \gls{FDA} criterion would typically be determined by self-reported race.
Across the \qty{98}{patients} used for calibration the reported categories are Asian (\num{27}), Caucasian (\num{25}), African American (\num{24}), and Hispanic (\num{7}), with \num{12} spread over multi-ethnic combinations and \num{3} not reported; \num{26} of them fall in the two categories treated here as darkly pigmented (i.e., African American and Other/Multiethnic African American).
Race is a coarse and poor stand-in for skin pigmentation, and it is used only to replicate what the \gls{FDA} guidance suggests.
\par

\subsection*{Calibration accuracy and skin-tone dependence}
\subsubsection*{Three calibration cases}
\begin{figure}[!htb]
	\glsreset{SaO2}\glsreset{ABG}\glsreset{RoR}\glsreset{mBLL}\glsreset{M}
	\begin{center}
		\includegraphics{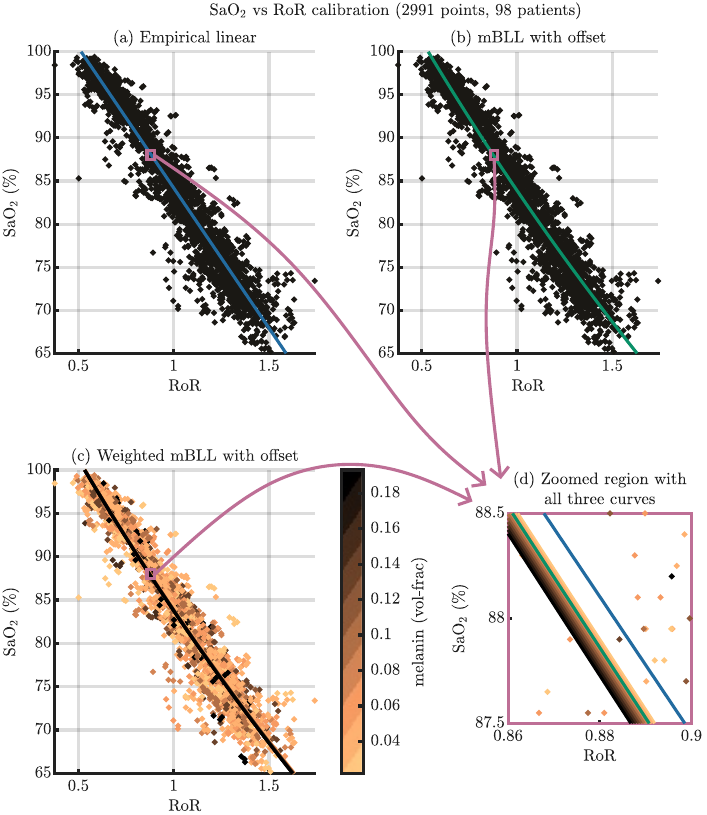}
	\end{center}
	\caption{
		Qualitative comparison of the three calibration cases.
		Plots show \gls{SaO2} from \gls{ABG} analysis versus the measured \gls{RoR} for \num{98} patients and \num{2991} blood-draws.
		Curves in these plots are fitted to this entire dataset.
		(a)~Empirical linear calibration (Equation~\ref{equ:linCal}) recovering $a=\qty{-32.53}{\percent}$ and $b=\qty{116.73}{\percent}$ in \gls{SaO2} units.
		(b)~\Gls{mBLL}-derived calibration with offset (Equation~\ref{equ:mBLLcal}) recovering path length ratio $\Gamma=\num{0.987}$ and offset $o=\qty{10.21}{\percent}$ in \gls{SaO2}.
		(c)~Detected-spectrum-weighted \gls{mBLL}-derived calibration with offset accounting for the spectral coloring of \gls{M} (Equation~\ref{equ:WmBLLcal}) recovering $\Gamma=\num{0.957}$ and $o=\qty{9.85}{\percent}$ in \gls{SaO2}. This case gives a curve that depends on \gls{M}.
		(d)~Zoomed region shown as a pink box in (a)--(c) showing the slight difference among the three curves and the differing \gls{M}-dependent curves for case (c).
	}
	\label{fig:cases}
\end{figure}

The first calibration case considered in this work is the conventional empirical linear (Equation~\ref{equ:linCal}); case two considers a calibration model based on \gls{mBLL} assuming monochromatic light sources\cite{Blaney_JBO24_CriticalAnalysis} (Equation~\ref{equ:mBLLcal}); and case three is that same recovery with detected-spectrum-weighted coefficients which account for spectral coloring of polychromatic sources\cite{Blaney_JBO26_BroadlinewidthSources} (Equation~\ref{equ:WmBLLcal}).
Cases one and two are blind to melanin---neither takes \gls{M} as an input.
Case three is the only melanin-aware case, since its detected spectrum, and hence its weighted extinction coefficients and weighted path-length ratio, depends on \gls{M}.
Case three also requires either knowledge or an assumption of the sources' bandwidth; in this work we assumed a red bandwidth of \qty{25}{\nano\meter} and an \gls{IR} bandwidth twice that (i.e., \qty{50}{\nano\meter}).
All three cases have two \gls{DOF} and differ only in form.
\par

Fitted over the whole calibration set of \qty{98}{patients}, case one gave $a=\qty{-32.53}{\percent}$ with $b=\qty{116.73}{\percent}$ both in \gls{SaO2} units (Equation~\ref{equ:linCal}), case two gave path-length ratio $\Gamma=\num{0.987}$ with offset $o=\qty{10.21}{\percent}$ in \gls{SaO2} (Equation~\ref{equ:mBLLcal}), and case three $\Gamma=\num{0.957}$ with $o=\qty{9.85}{\percent}$ in \gls{SaO2} (Equation~\ref{equ:WmBLLcal}; Figure~\ref{fig:cases}).
Accuracy---considering the whole fit of \qty{98}{patients}---is nearly identical across the three cases, at $A_{rms}$ of \qtylist{2.420;2.416;2.418}{\percent} in \gls{SaO2} units, respectively.
Similar average accuracies were found in a \num{200}-fold cross-validation using \gls{FDA} complaint training sets\cite{FDACDRH_13_PulseOximeters}.
The median held-out $A_{rms}$s for the three cases were \qtylist{2.494;2.482;2.484}{\percent} in \gls{SaO2}, respectively (Figure~\ref{fig:crossval}(a)).
Every held-out fold of every case in the \num{200}-fold cross-validation fell below the \qty{3.0}{\percent} transmittance limit set by the \gls{FDA} guidance\cite{FDACDRH_13_PulseOximeters}.
Therefore, it is reasonable to conclude that these three calibration cases perform practically identically in terms of overall accuracy. 
\par

\subsubsection*{Skin-tone dependence}
\begin{figure}[!htb]
	\glsreset{FDA}\glsreset{SaO2}\glsreset{ABG}\glsreset{M}\glsreset{ITA}
	\begin{center}
		\includegraphics{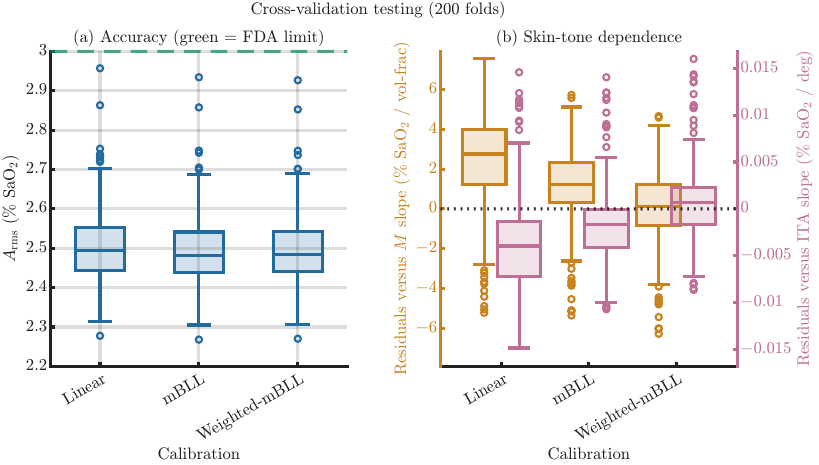}
	\end{center}
	\caption{
		Quantitative comparison of the three calibration cases using \num{200}-fold cross-validation across \num{98} patients.
		Training sets consisted of \num{10} patients that satisfied the 2013 \gls{FDA} guidance\cite{FDACDRH_13_PulseOximeters}; the remaining \num{88} patients formed the testing set.
		Plots present testing set metrics.
		(a)~Root-mean-square accuracy in \gls{SaO2} including a green dashed line at the \gls{FDA} acceptance limit.
		(b)~Skin-tone or -color dependence of the calibration expressed as the slope of the residuals between recovered and \gls{ABG} \gls{SaO2} versus either \gls{M} (orange) or \gls{ITA} (pink).
		A slope of \num{0} (dotted black line) indicates no \gls{M} or \gls{ITA} dependence.
	}
	\label{fig:crossval}
\end{figure}

We measured skin-tone or -color dependence in terms of the linear slope of the \gls{SaO2} residual---the difference between the recovered value and the gold standard \gls{ABG} value---versus either \gls{M} or \gls{ITA}.
These results are shown in Figure~\ref{fig:crossval}(b) for the \num{200} cross-validation folds. 
Considering \gls{M} as the skin-tone variable, the median held-out slope was $+\num{2.76}$ percentage points of \gls{SaO2} per unit volume fraction for case one, $+\num{1.24}$ for case two, and $+\num{0.10}$ for case three.
Against \gls{ITA} the same pattern appears with the opposite sign---since \gls{ITA} is more negative with higher \gls{M} (Figure~\ref{fig:spectra_ITA_M}(b))---at median held-out slopes of $\num{-0.0040}$, $\num{-0.0017}$, and $+\num{0.0007}$ percentage points of \gls{SaO2} per degree \gls{ITA}.
The ordering above therefore indicates the direction in which a spectral-coloring correction moves the residual.
However, we cannot conclude that a melanin-dependent bias has been measured in these data, since the fold-to-fold range of each slope contains zero.
Overall, we report the correction as applied and physically motivated while making no claim that it removes a measured skin-tone bias in these data.
\par

\section*{Discussion}
This work considered three pulse oximetry calibration cases which were fit to the same \num{2991} blood-draws from \qty{98}{patients}.
The cases differ only in the function relating \gls{SaO2} to \gls{RoR}; importantly each contains two \acrfull{DOF}.
They are indistinguishable in accuracy, differing by less than \qty{0.02}{\percent} in \gls{SaO2} units considering root-mean-square error.
These calibration cases differ in what each form is able to represent, and how much the residuals depend on skin tone.
\par

The first calibration case (Equation~\ref{equ:linCal} and Figure~\ref{fig:cases}(a)) is entirely empirical.
Its linear equation's slope and intercept were not derived from physical principles; however, these parameters can be expressed in terms of physical quantities if the \gls{mBLL}-derived second case is linearized\cite{Blaney_JBO24_CriticalAnalysis}.
The \gls{mBLL}-derived form shows \gls{SaO2} as a ratio of two linear expressions with \gls{RoR} (i.e., linear-fractional: $y=(ax+b)/(cx+d)$).
However, the linear-fractional expression only contains one unknown \gls{DOF} (i.e., the ratio of path lengths $\Gamma$) with other coefficients being combinations of extinction coefficients which are fixed if the optical wavelengths are known\cite{Blaney_JBO24_CriticalAnalysis,Blaney_JBO26_BroadlinewidthSources}.
Calibration cases two and three (Equations~\ref{equ:mBLLcal}\&\ref{equ:WmBLLcal} and Figure~\ref{fig:cases}(b)\&(c)) consider a combination of the theoretical \gls{mBLL} expression plus an additive empirical constant (i.e., the offset $o$).
This offset parameter was adopted for two reasons: first, to match the number of \gls{DOF} in case one; second, to substantially improve fit quality as a single \gls{DOF} fit was unable to adequately capture the \gls{SaO2} versus \gls{RoR} relationship in the experimental data.
The \gls{mBLL}-derived linear-fractional form generates a curved line---with or without the offset parameter---which a simple linear equation with the same number of \gls{DOF} cannot recreate.
Case three (Equation~\ref{equ:WmBLLcal} and Figure~\ref{fig:cases}(c)) adds no additional \gls{DOF}, but considers spectral coloring of broadband sources by tissue including \gls{M}\cite{Blaney_JBO26_BroadlinewidthSources}.
This makes \gls{M} an independent variable and turns the single curve into a family of curves dependent on skin tone.
Therefore, case three would need a measurement of \gls{M} and knowledge of the light source bandwidth to implement in practice.
All three cases carry two \gls{DOF}, but only cases two and three are physically derived and interpretable.
\par

The curvature of cases two and three is small but not negligible over the range of \gls{SaO2} considered.
Across the calibration set's \gls{RoR} range of \numrange{0.376}{1.740}, the case-two curve departs from the best straight line through it by up to \qty{1.36}{\percent} in \gls{SaO2} and the case-three curve by up to \qty{1.39}{\percent} in \gls{SaO2}.
This curvature results in a fit curve offset from the linear case despite having the same number of \gls{DOF} (Figure~\ref{fig:cases}(d)) and may have implications for values recovered outside the calibration range---where extrapolation relies solely on the model's form since no data was present in the fit.
Despite the different curves recovered by the fit for the three cases, $A_{rms}$ is almost unaffected (Figure~\ref{fig:crossval}(a)) because the differences in the curves are largest where there is the least experimental data---only \qty{5.8}{\percent} of blood-draws lie above a \gls{RoR} of \num{1.4} and \qty{5.2}{\percent} below \num{0.6}---so the three curves separate mainly outside the crowded middle of the distribution that dominates $A_{rms}$.
Since there is no observable difference in accuracy of the three calibration cases, we cannot recommend one over another on that basis, but we do expect a difference in measurements which extend below the calibrated \gls{SaO2} range where the shape of the extrapolated curve matters\cite{Wu_JBO23_SelfcalibratedPulse,Feiner_AA07_DarkSkin}.
We would expect cases two and three to perform better in this scenario; however, we have no evidence of this in these data, since extremely low \glspl{SaO2} were not reached.
\par

The \gls{mBLL}-based calibration cases' fitted path-length ratios (i.e., $\Gamma$s) recover values above the modeled monochromatic value of $\Gamma_{mono}=\num{0.850}$ considering the total path-lengths in a bulk tissue slab as modeled in Ref.~\citenum{Blaney_JBO26_BroadlinewidthSources}.
However, agreement was not expected, since we know considering $\Gamma$ in that way is, strictly speaking, not correct and only gives a rough estimate of the true value\cite{Blaney_JBO24_CriticalAnalysis}. 
The \gls{mBLL} expression underlying Equations~\ref{equ:mBLLcal}\&\ref{equ:WmBLLcal} contains a ratio of mean \emph{partial} optical path-lengths in the \emph{pulsatile arterial volume}---the tissue regions the arterioles expand into---and not a ratio of total optical path-lengths through the entire bulk tissue.
A preliminary Monte Carlo study of these partial path-lengths supports that reading, indicating that the relevant quantity is the path-length in the annulus surrounding the blood vessels rather than the bulk path-length, and that this ratio itself varies with \gls{SaO2}\cite{Blaney_ProcSPIE25_PreliminaryInvestigation}.
At \num{0.987} for case two and \num{0.957} for case three they are much closer to unity compared to the bulk value, which is consistent with the pulsatile partial paths at the two wavelengths being nearly equal.
\par

Interpretation of the additive offset (i.e., $o$) is more difficult since it is not arrived at by the \gls{mBLL} derivation.
Cases two and three recovered \qty{10.21}{\percent} and \qty{9.85}{\percent} for the offset, respectively.
An offset of that magnitude is unlikely due to rounding or noise---it says the \gls{mBLL} expression models roughly \qty{10}{\percent} in \gls{SaO2} low across the whole cohort---and it is stable across the cross-validation folds, suggesting it is systematic.
Candidate explanations include venous, tissue, or water contamination of the pulsatile signal\cite{Walton_JCMC10_MeasuringVenous,Kainerstorfer_JBO16_OpticalOximetry} or an offset in the \gls{ABG} values (e.g., the dataset contains \gls{SaO2} from \gls{ABG} above \qty{100}{\percent}), though the latter is unlikely given the magnitude of the offset.
To interpret this offset one would need to carefully consider which assumptions in Ref.~\citenum{Blaney_JBO24_CriticalAnalysis} are broken in the real-world measurement.
Distinguishing among these explanations is beyond the scope of this work, therefore in this context we consider $o$ an uninterpreted empirical parameter similar to $b$ in the linear case.
\par

Moving to the three calibration cases' skin-tone dependence, we can discuss the interpretation of Figure~\ref{fig:crossval}(b).
The ordering of the calibration cases shows what each case is able to represent: the empirical linear case shows the most uncorrected skin-tone dependence, the nominal \gls{mBLL}-derived recovery---which considers monochromatic wavelengths thus no source linewidth---shows less dependence, and case three, which accounts for spectral coloring of a finite linewidth given \gls{M}, sits near zero skin-tone or -color dependence.
An important caveat is that none of these slopes are resolved by these data (e.g., every case's \qty{95}{\percent} fold-to-fold spread interval contains zero dependence).
We therefore report an observed change in dependence across three ways of fitting the same data, and not a measured skin-tone bias in case one nor a statistically demonstrated removal of a skin-tone bias.
The expected cohort size required to resolve such a skin-tone dependence in the residuals is therefore a useful point of comparison.
This calibration set holds \qty{98}{patients}, whereas the clinical reports of skin-tone-dependent pulse-oximeter error utilize far larger cohorts---\qty{1565}{patients} in one and \qty{8392}{patients} in another cohort, both from Ref.~\citenum{SjodingMichaelW._NEJM20_RacialBias}.
Therefore, a study aiming to resolve a significant slope of \gls{SaO2} residual versus \gls{M} is expected to need on the order of ten times the patients here.
So the absence of a resolved dependence in these data does not negate the change we observe between the calibration cases, but points to the need for a larger cohort study.
Public cohorts of the required size are becoming available---for example, the Blood-gas and Oximetry Linked Dataset (BOLD) combines \num{49099} paired measurements from three intensive-care databases\cite{Matos_SciData24_BOLDBloodgas}---though such resources record paired commercial oximeter and \gls{ABG} recovered values rather than the raw \gls{PPG} traces---or \gls{RoR}---and skin reflectance spectra that the correction tested here requires.
Nonetheless, our interpretation of these results shows promise for calibration cases two and three, especially three which may correct for the spectral-coloring portion of the skin-tone bias in pulse oximetry.
\par

In conclusion, this work proposes and tests two \gls{mBLL}-derived pulse-oximetry calibration methods (i.e., cases two and three).
Although case three's correction was suggested in Ref.~\citenum{Blaney_JBO26_BroadlinewidthSources}, this is the first work with sufficient detail to implement it---see the Methods section---and test it on experimental data.
While case three is the best-performing case, it is also the most complex to implement due to its requirement of a separate measurement of \gls{M} for each patient.
This suggests that case two is the most feasible to implement for current oximeter technology; however, we must note that case two is melanin-blind and only case three is melanin-aware.
Furthermore, since we propose calibration methods, the melanin and linewidth corrections cannot be directly applied to current pulse oximeters.
Future work may utilize the theory and model presented here to derive such an oximeter correction.
Derivation of such a correction may be necessary to reach a cohort size with enough statistical power to show a statistically significant skin-tone bias and its correction, since datasets of such size exist for clinical commercial oximeter readings but not for controlled desaturation studies.
As low-hanging fruit, in future desaturation-based pulse-oximeter calibration studies we suggest that \gls{mBLL}-based calibration models be considered, especially case two which does not require a separate skin pigmentation measurement and could easily replace linear calibration since it contains the same number of \gls{DOF}.
\par

\section*{Methods}
The analysis in this work can be reproduced using the code in the accompanying public GitHub repository\cite{Blaney_26_MelaninLinewidthcorrected} and the data that are publicly available in the OpenOximetry Repository v1.1.1 from PhysioNet\cite{Fong_24_OpenOximetryRepository}.
These data were collected under the UCSF Hypoxia Lab controlled-desaturation protocol, aligned with ISO 80601-2-61, in which healthy adult subjects breathe titrated nitrogen, oxygen, and carbon dioxide gas mixtures to hold a sequence of stable arterial oxygen saturation plateaus spanning about \qtyrange{70}{100}{\percent}.
At each plateau, arterial blood drawn from a radial-artery catheter was analyzed by \gls{ABG} oximetry (Radiometer ABL90 Flex Plus) to yield a gold-standard \gls{SaO2} measurement, which was recorded simultaneously with a custom pulse-oximeter reporting raw \gls{PPG} (Analog Devices MAX86171).
Skin pigmentation was quantified with a reflectance spectrophotometer designed for color measurement (Konica Minolta CM-700d).
No human-subject data other than the OpenOximetry repository were collected or used in this work.
\par

The OpenOximetry study was approved by the UCSF IRB \#21-35637 and \#23-40212---the authors of this work were not associated with or involved in the study.
Informed consent, including consent to share de-identified data, was obtained from all participants by the original study investigators, and all data included in the repository were collected in accordance with the ethical principles of the Declaration of Helsinki\cite{Fong_24_OpenOximetryRepository}.
Large language models were used for proofreading and editing of the work; they were not used to generate or interpret results, and the authors take full responsibility for the content of this work.
\par

\subsection*{Indexing \gls{ABG} data}
The first stage of the analysis indexed the OpenOximetry\cite{Fong_24_OpenOximetryRepository} repository to extract \gls{SaO2} data from the \gls{ABG} measurements.
From the blood-gas table the patient identifiers, encounter identifiers, sample numbers, and \gls{SaO2} measurements were extracted---rows missing one of these values were excluded.
Rows in the blood-gas table were also excluded if $\text{\acrshort{SaO2}} > \qty{100}{\percent}$ or if their encounter and patient identifiers did not appear together as a registered pair in the encounter table.
For each sample the median of repeated \gls{ABG} draws was considered.
In total, the blood-gas table included \qty{32877}{rows} associated with \qty{224}{patients} which were processed to result in \qty{17756}{samples} of \gls{SaO2} from \qty{218}{patients} (Figure~\ref{fig:cohort} -- blue set).
This step of the analysis is found in the \texttt{A\_loadSaO2\_openox.m} script in the GitHub repository\cite{Blaney_26_MelaninLinewidthcorrected}.
\par

\subsection*{Ratio-of-ratios extraction from raw \gls{PPG}}
Each encounter's raw \gls{PPG} data was read using the PhysioNet WaveForm DataBase (WFDB) Toolbox\cite{Silva_JORS14_OpensourceToolbox}.
These consisted of \qty{86}{\hertz} recordings at \qtylist{660;910}{\nano\meter} (i.e., red and \gls{IR}) optical wavelengths.
Missing samples were linearly interpolated with nearest-value end filling.
Then, spike or step artifacts were detected and their indexes saved (Algorithm~\ref{alg:artifact}).
The red and \gls{IR} artifact flags were combined by logical union into one mask, and only contiguous valid segments of at least \qty{3}{\second} were considered.
\par

\begin{algorithm}[htb]
	\caption{Spike and step artifact detection}
	\label{alg:artifact}
	\begin{algorithmic}[1]
		\Require Full-encounter \gls{PPG}~trace~$I$; sampling~rate~$f_s$;
		spike~window~$w_{sp}\gets\qty{2}{\second}$;
		spike~threshold~$k_{sp}\gets5$;
		step~half-window~$w_{st}\gets\qty{0.25}{\second}$;
		step~threshold~$k_{st}\gets6$;
		step~floor~$m_{st}\gets0.0015$;
		min~valid~segment~length~$\Delta t_{min}\gets\qty{3}{\second}$
		
		\Ensure Valid artifact-free segments $\{s\}$
		
		\State moving baseline $B\gets$ moving median of $I$ over $\max(3,\operatorname{round}(w_{sp}f_s))$ samples
		\State relative change $X\gets(I-B)/B$
		\State pseudo standard deviation $\sigma_X\gets\operatorname{median}(|X-\operatorname{median}(X)|)/\operatorname{norminv}(0.75)$
		\State spike mask $A_{sp}\gets(|X|\geq k_{sp}\sigma_X)$
		
		\State trailing baseline $B^-\gets$ moving median of $I$ over $\max(2,\operatorname{round}(w_{st}f_s))$ samples before the sample
		\State leading baseline $B^+\gets$ moving median of $I$ over $\max(2,\operatorname{round}(w_{st}f_s))$ samples after the sample
		\State relative edge change $Y\gets |B^- - B^+|/B$
		\State pseudo edge standard deviation $\sigma_Y\gets\operatorname{median}(|Y-\operatorname{median}(Y)|)/\operatorname{norminv}(0.75)$
		\State step mask $A_{st}\gets(Y\geq \max(m_{st}, k_{st}\sigma_Y))$
		
		\State artifact trace $A\gets A_{sp} \cup A_{st}$
		\State valid segments $\{s\}\gets$ all segments between artifacts in $A$ longer than $\Delta t_{min}$
		
		\State \Return $\{s\}$
	\end{algorithmic}
\end{algorithm}

The \gls{RoR} was extracted on two types of fixed-length analysis windows: 1) \qty{20}{\second} windows swept in \qty{10}{\second} steps to form a \gls{RoR} trace for synchronization with \gls{ABG}, and 2) \qty{30}{\second} windows centered on each \gls{ABG} blood-draw to yield a \gls{RoR} for each \gls{SaO2} measurement.
Within each analysis window, the \gls{HR} was estimated as the dominant Fourier-domain magnitude peak in the range \qtyrange{0.8}{3.5}{\hertz} considering the \gls{PPG} signal for the \gls{IR} optical wavelength.
Per-beat \gls{RoR} was estimated based on the peak-to-peak amplitude of the raw \gls{PPG} (Algorithm~\ref{alg:ppbeats}).
A window's \gls{RoR} was the median of its per-beat values after rejecting beat-level outliers beyond \num{3} pseudo standard deviations (i.e., $3\times\operatorname{median}(|X-\operatorname{median}(X)|)/\operatorname{norminv}(0.75)$) of the median.
A window was accepted only when it retained at least \num{20} beats, its pseudo standard deviation normalized by its median was at most \num{0.08}, and the number of artifact samples was less than \num{0.3} times the window length.
These criteria are deliberately \emph{not} applied to the sliding windows of the synchronization trace (Algorithm~\ref{alg:sync}).
\par

\begin{algorithm}[htb]
	\caption{Peak-to-peak \gls{RoR} estimator}
	\label{alg:ppbeats}
	\begin{algorithmic}[1]
		\Require Windowed red/\gls{IR}~\gls{PPG}~traces~$I_r,I_i$;
		valid~segments~$\{s\}$;
		heart~rate~$f_{HR}$; 
		sampling~rate~$f_s$;
		baseline~span~$w_b\gets3$~periods;
		min~trough~spacing~$w_t\gets0.5$~periods;
		min~trough~prominence~$p\gets0.2$
		
		\Ensure Per-beat \gls{RoR}
		
		\State baseline window $W\gets\max(3,\operatorname{round}(w_b f_s/f_{HR}))$ samples
		\State de-trended average $I^\prime \gets [(I_r-\operatorname{movmedian}(I_r,W))+(I_i-\operatorname{movmedian}(I_i,W))]/2$
		
		\For{each segment $s$}
		\State troughs $T\gets$ minima of $I^\prime$ on $s$, spaced $\ge w_t$ periods, prominence $\ge p\times(Q_{0.95}-Q_{0.05})$
		\For{each period between consecutive troughs in $T$}
		\State $AC_y\gets\max (I_y)-\min (I_y)$, with $y\in\{r,i\}$
		\State $DC_y\gets(\max (I_y)+\min (I_y))/2$, with $y\in\{r,i\}$
		\If{$AC_r,AC_i,DC_r,DC_i>0$}
		\State append $(AC_r/DC_r)/(AC_i/DC_i)$ to \gls{RoR}
		\EndIf
		\EndFor
		\EndFor
		\State \Return \gls{RoR}
	\end{algorithmic}
\end{algorithm}

Synchronization was required since the raw \gls{PPG} and the monitor files containing \gls{ABG} blood-draw times did not share a time axis, so the offset was determined from waveform shape using Algorithm~\ref{alg:sync}.
This was achieved by finding the temporal offset corresponding to the maximum cross-correlation between the \gls{RoR} synchronization trace and the negative of the reported \gls{SpO2} from a commercial oximeter.
The synchronization quality was determined using the Pearson correlation between these two signals at that temporal offset, and encounters with Pearson correlation coefficient below \num{0.9} were excluded.
\par

This stage of the analysis processed the \qty{139}{patients} associated with an encounter carrying a raw \gls{PPG} record (Figure~\ref{fig:cohort} -- green set).
This totaled \qty{7723}{samples} from candidate blood-draws, of which \qty{3376}{samples} were associated with \gls{RoR} values which passed quality thresholds, resulting in that number of \gls{RoR} samples for \qty{106}{patients} (Figure~\ref{fig:cohort} -- orange set bar).
The step of the analysis described in this section can be found in the \texttt{B\_extractRoR\_openox.m} script in the GitHub repository\cite{Blaney_26_MelaninLinewidthcorrected}.
\par

\begin{algorithm}[htb]
	\caption{\gls{PPG} \gls{RoR} and \gls{ABG} blood-draw time synchronization}
	\label{alg:sync}
	\begin{algorithmic}[1]
		\Require Full-encounter red/\gls{IR}~\gls{PPG}~traces~$I_r,I_i$;
		reference~\gls{SpO2}~trace~$S$~(in~\gls{ABG}~time);
		\gls{ABG}~blood-draw~times~$\{D\}$;
		sync~window~$w_s\gets\qty{20}{\second}$;
		sync~step~$\Delta t_s\gets\qty{10}{\second}$;
		resample~grid~$\delta\gets\qty{1}{\second}$;
		quality~threshold~$r_{\min}\gets\num{0.9}$;
		blood-draw~half-window~$h\gets\qty{15}{\second}$
		
		\Ensure \Gls{RoR} for each \gls{ABG} blood-draw
		
		\State slow \gls{RoR} trace \acrshort{RoR}$^\prime\gets$ median of the windowed peak-to-peak estimator (Alg.~\ref{alg:ppbeats}) on $w_s$ windows spaced by $\Delta t_s$, requiring that the window contains at least two beats
		
		\State monitor start $t_0\gets\min(\text{time of }S)$
		\State \acrshort{RoR}$^\prime_{z},S_z\gets$ $z$-scores of \acrshort{RoR}$^\prime_{pp}$ and $-S$, each resampled to $\delta$ spacing on their own time axis 
		
		\State lag $\Delta\gets$ lag time of $\max(\operatorname{xcorr}(\acrshort{RoR}^\prime_{pp,z},S_z))$
		\State synced $r\gets$ Pearson correlation coefficient of $\acrshort{RoR}^\prime_{pp,z}$ and $S_z$ after shifting $S_z$ by $\Delta$
		
		\If{$r<r_{\min}$}
		\State \Return exclude the encounter
		\EndIf
		
		\State temporal sync offset $\tau\gets t_0-\Delta$, which maps monitor time onto \gls{PPG}-record time
		
		\For{blood-draw time $D$}
		\State compute \gls{RoR} by Alg.~\ref{alg:ppbeats} on the \(\pm h\) window at $D-\tau$
		\EndFor
		
		\State\Return \gls{RoR}
	\end{algorithmic}
\end{algorithm}

\subsection*{Skin tone and melanin fitting}
\subsubsection*{Skin optics model}
The analysis in this work focuses on \gls{M} because melanin is the optical absorber that models different skin tones directly\cite{Vasudevan_Comm.Med.24_MelanometryObjective,Kollias_ClinicsinDermatology95_PhysicalBasis}.
\Gls{M} was fit for using the measured \qtyrange{400}{700}{\nano\meter} reflectance spectra.
The fit modeled the epidermis as a transmissive filter with effective optical thickness of \qty{0.2}{\milli\meter} on top of a semi-infinite medium representing the dermis similar to previous work\cite{Blaney_JBO26_BroadlinewidthSources,Yudovsky_J.Biophotonics11_RetrievingSkin}.
This model determined the \gls{mua} of the epidermis solely from the \gls{M} absorption, which has been reported as follows\cite{Jacques_PMB13_OpticalProperties}:
\begin{equation}\label{equ:muaEpi}
	{\acrshort{mua}}_{,epi}(\lambda) = 
	\acrshort{M} \times  \qty{51.9}{\per\milli\meter}\left(\frac{\lambda}{\qty{500}{\nano\meter}}\right)^{-3.5}
\end{equation}
where \qty{51.9}{\per\milli\meter} represents the \gls{mua} for $\acrshort{M}=1$ at \qty{500}{\nano\meter} and the \num{-3.5} power represents the wavelength dependence.
For the dermis, model determined the \gls{mua} from the \gls{HbT} concentration, \gls{StO2}, and \gls{W} as follows\cite{Blaney_JBO24_DualratioApproach}:
\begin{equation}\label{equ:muaDer}
	{\acrshort{mua}}_{,der}(\lambda) = 
	\acrshort{StO2} [\acrshort{HbT}] \epsilon_{\text{HbO\textsubscript{2}}}(\lambda) + 
	(1-\acrshort{StO2}) [\acrshort{HbT}] \epsilon_{\text{Hb}}(\lambda) + 
	\acrshort{W} {\acrshort{mua}}_{,water}(\lambda)
\end{equation}
where $\epsilon$'s represent Molar extinction coefficients of either \gls{HbO2} or \gls{Hb} according to the subscript.
The \gls{musp} of the dermis was modeled as a combination of Rayleigh and Mie scattering as follows\cite{Jacques_PMB13_OpticalProperties,Blaney_JBO24_DualratioApproach}:
\begin{equation}\label{equ:muspDer}
	{\acrshort{musp}}_{,der}(\lambda) = 
	{\acrshort{musp}}_{,der}(\qty{500}{\nano\meter})
	\left( 
	f_{Ray} \left(\frac{\lambda}{\qty{500}{\nano\meter}}\right)^{-4} + 
	(1-f_{Ray}) \left(\frac{\lambda}{\qty{500}{\nano\meter}}\right)^{-b_{Mie}}
	\right)
\end{equation}
where $f_{Ray}$ is the fraction of Rayleigh scattering and $b_{Mie}$ is the Mie scattering power.
\par

Spectra were acquired using a Konica Minolta CM-700d on the small-aperture setting---three repeated measurements per site---under a D65 / \ang{2} illuminant--observer setting\cite{UCSFHypoxiaLab_25_ProtocolSkin}; in that configuration the instrument diffusely illuminates a \qty{6}{\milli\meter} diameter port and views a \qty{3}{\milli\meter} diameter measurement area at \ang{8} from the normal\cite{KonicaMinolta_18_SpectrophotometerCM700d}.
The illuminated port and the collection area are coaxial disks of \emph{different} radii, $r_{ill}=\qty{3}{\milli\meter}$ and $r_{col}=\qty{1.5}{\milli\meter}$, so the \glspl{rho} that contribute to the measurement are distributed following the overlap area ($\Lambda$) of those two disks with centers separated by \gls{rho}:
\begin{equation}
	\Lambda(\rho) =
	\begin{cases}
		\pi r_{col}^2
		& \rho \leq r_{ill}-r_{col}\\[8pt]
		\begin{aligned}
			&r_{col}^2\cos^{-1}\left(\frac{\rho^2+r_{col}^2-r_{ill}^2}{2\rho r_{col}}\right)
			+ r_{ill}^2\cos^{-1}\left(\frac{\rho^2+r_{ill}^2-r_{col}^2}{2\rho r_{ill}}\right)\\
			&\quad - \frac{1}{2}\sqrt{\left[(r_{ill}+r_{col})^2-\rho^2\right]
				\left[\rho^2-(r_{ill}-r_{col})^2\right]}
		\end{aligned}
		& r_{ill}-r_{col} < \rho < r_{ill}+r_{col}\\[8pt]
		0
		& \rho \geq r_{ill}+r_{col}
	\end{cases}
	\label{eq:overlap}
\end{equation}
Considering the dermis Green's function, $G_{der}(\rho,\lambda)$, for the continuous-wave diffuse reflectance\cite{Blaney_JIOHS24_SpatialSensitivity} from a semi-infinite medium at separation $\rho$, and the epidermal transmission, $T_{epi}(\lambda)$, through a non-scattering thin slab\cite{Blaney_JBO26_BroadlinewidthSources}, the total diffuse reflectance ($R_{skin}(\lambda)$) leaving the skin over the measurement aperture is (n.b., in this formulation, $R_{skin}(\lambda)$ is unit-less as it is normalized by the source power-spectral-density entering the boundary):
\begin{equation}
	R_{skin}(\lambda) = T_{epi}(\lambda)
	\int_0^{r_{ill}+r_{col}} G_{der}(\rho,\lambda)\,\frac{\Lambda(\rho)}{\Lambda(0)}\,2\pi\rho\,d\rho
\end{equation}
where $G_{der}(\rho,\lambda)$ depends on the dermal absorption (Equation~\ref{equ:muaDer}) and scattering (Equation~\ref{equ:muspDer}), and $T_{epi}(\lambda)$ depends on the epidermal absorption (Equation~\ref{equ:muaEpi}) and effective optical thickness $L_{epi}$:
\begin{equation}\label{equ:Tepi}
	T_{epi}(\lambda)=e^{-2L_{epi}\,{\acrshort{mua}}_{,epi}(\lambda)}
\end{equation}
(n.b., $L_{epi}$ is multiplied by two to account for the round trip distance).
Finally, the measured total diffuse reflectance ($R_{meas}(\lambda)$) considers the front-surface Saunderson specular reflection term ($K_{spec}$) in which a fraction $K_{spec}$ of the incident light is reflected at the air skin boundary\cite{Garcia-Valenzuela_J.Phys.:Conf.Ser.11_AssessmentSaunderson,Berns_19_BillmeyerSaltzmans}:
\begin{equation}\label{equ:specRef}
	R_{meas}(\lambda) = K_{spec} + \left(1-K_{spec}\right) R_{skin}(\lambda)
\end{equation}
(n.b., $R_{meas}(\lambda)$ is unit-less as it is normalized by the power-spectral-density of the source emission).
Internal re-reflection at that boundary is accounted for by the refractive-index mismatch considered in the $G_{der}(\rho,\lambda)$ term.
The fitting procedure minimized the squared difference between the modeled and observed $R_{meas}(\lambda)$.
\par

\subsubsection*{Spectra pre-processing and fitting}
The fit was restricted to spectra collected on the finger dorsal (top) side, palmar (bottom) side, forehead, and inner upper arm measurement sites\cite{UCSFHypoxiaLab_25_ProtocolSkin}.
Each spectrum sampled \qtyrange{400}{700}{\nano\meter} every \qty{10}{\nano\meter} resulting in \num{31} individual wavelengths.
A spectrum was rejected if any of its \num{31} reflectance values was non-positive, or if its peak reflectance fell outside $[\num{0.01}, \num{0.90})$.
Of the \num{9517} individual spectra at the four fitted sites, \num{288} were incomplete and \num{1} carried a non-positive value, leaving \num{9228}.
Surviving spectra were grouped by patient and site, and a group was carried forward only if it held at least \num{3} replicates; \num{2} groups held only \num{2} replicates each and were excluded, removing \num{4} spectra and leaving \num{9224} in \num{810} groups (i.e., patient-site combos).
A replicate was discarded when its level departed from the group median by more than both \num{5} pseudo standard deviations and \num{0.25} in $\log_{10}$ units; the group itself was retained and pooled from the replicates that survived---the pseudo standard deviation again being $\operatorname{median}(|X-\operatorname{median}(X)|)/\operatorname{norminv}(0.75)$ over the levels in that group.
This discarded \num{7} replicates drawn from \num{5} groups, and the remaining \num{9217} spectra were averaged wavelength by wavelength into one spectrum per patient-and-site combination (i.e., group), giving \num{810} averaged spectra.

Water volume fraction (\num{0.65}), dry-tissue refractive index (\num{1.514}), reduced-scattering amplitude (\(\mu_s'(\qty{500}{\nano\meter}) = \qty{4.36}{\per\milli\meter}\)), and Mie scattering power (\(b = \num{0.562}\)) were fixed at dermis literature values\cite{Blaney_JBO24_DualratioApproach,Jacques_PMB13_OpticalProperties}.
Each remaining optical-property parameter was fit for, either with a unique value for each spectrum or a shared value across all spectra.
Four population-shared parameters (i.e., shared values across all spectra) were considered, namely the three dermis properties \gls{HbT}, \gls{StO2} (Equation~\ref{equ:muaDer}), and the Rayleigh scattering fraction (Equation~\ref{equ:muspDer}), together with the front-surface specular reflectance factor (Equation~\ref{equ:specRef}).
The parameter of interest, \gls{M} (Equation~\ref{equ:muaEpi}), found a unique value for each average spectrum.
Any individual spectrum whose coefficient of determination---referenced to that spectrum's own mean---fell below \num{0.50} was discarded and the fit rerun until all remaining spectra met this threshold; \num{19} of the \num{810} average spectra were discarded by this threshold.
\par

\subsubsection*{Color quantification}
Skin pigmentation was also quantified in terms of color using the \gls{ITA}.
\Gls{ITA} was computed from CIE~$L^*a^*b^*$ coordinates ($L^*$ -- lightness; $a^*$ -- red/green; $b^*$ -- yellow/blue) derived from the measured reflectance spectrum---integrated against the CIE 1931 \ang{2} color-matching functions under D65 illumination\cite{Berns_19_BillmeyerSaltzmans}---as:
\begin{equation}
	\mathrm{\acrshort{ITA}}=\tan^{-1}\left(\frac{L^*-50}{b^*}\right)
\end{equation}
with \gls{ITA} expressed in degrees, and larger \gls{ITA} corresponding to lighter skin.
Despite this colorimetric descriptor not directly representing melanin, it was retained because it is simple and directly comparable to the repository's own colorimetry.
\par

\subsubsection*{Skin tone summary}
In total, this stage of the analysis processed \qty{810}{spectra} from \qty{221}{patients} (Figure~\ref{fig:cohort} -- purple set) each representing a unique patient and site combination for any of the four fitted sites.
Of these, \qty{791}{spectra} satisfied the goodness-of-fit requirement and finger \gls{M} values for \qty{196}{patients} were recovered (Figure~\ref{fig:cohort} -- brown set).
This entire step of the analysis is found in the \texttt{C\_extractM\_openox.m} script in the GitHub repository\cite{Blaney_26_MelaninLinewidthcorrected}.
\par

\subsection*{\Gls{SaO2} versus \gls{RoR} calibration}
A blood-draw data point (i.e., a \gls{SaO2} and \gls{RoR} pair) was used for calibration when it passed all quality thresholds and was matched with a \gls{M} estimate---this criterion was met by \num{2991} blood-draws from \qty{98}{patients} (Figure~\ref{fig:cohort} -- red intersection).
Three \gls{SaO2} versus \gls{RoR} models were considered for calibration.
The first case utilized the conventional empirical linear calibration:
\begin{equation}\label{equ:linCal}
	\acrshort{SaO2}(\acrshort{RoR}) = a \times \acrshort{RoR} + b
\end{equation}
where $a$ and $b$ are empirical slope and intercept free parameters, respectively.
The second case considered the \gls{mBLL}-derived expression\cite{Blaney_JBO24_CriticalAnalysis,Blaney_JBO26_BroadlinewidthSources} with an empirical additive offset ($o$) included to improve fit quality and match the number of \gls{DOF} in the empirical linear case:
\begin{equation}\label{equ:mBLLcal}
	\acrshort{SaO2}(\acrshort{RoR}) = \frac
	{-\epsilon_{\acrshort{Hb},r} + \epsilon_{\acrshort{Hb},i} \times \Gamma \times \acrshort{RoR}}
	{(\epsilon_{\acrshort{HbO2},r}-\epsilon_{\acrshort{Hb},r}) +  (\epsilon_{\acrshort{Hb},i}-\epsilon_{\acrshort{HbO2},i}) \times \Gamma \times \acrshort{RoR}} + 
	o
\end{equation}
where the $\epsilon$'s are the molar extinction coefficients for either \gls{HbO2} or \gls{Hb} according to the subscript, and either red or \gls{IR} wavelengths---$r$ or $i$ subscripts, respectively---and $\Gamma$ is the ratio of average optical path-lengths:
\begin{equation}
	\Gamma=\frac{\langle\ell\rangle_i}{\langle\ell\rangle_r}
\end{equation}
at \gls{IR} and red wavelengths.
The third case utilized \gls{mBLL} the same way as Equation~\ref{equ:mBLLcal}, but accounted for polychromatic sources, whereas Equation~\ref{equ:mBLLcal} assumed monochromatic light sources.
Polychromatic sources with \gls{FWHM} $w$ were considered by using detected spectrum weighted extinction coefficients ($\bar{\epsilon}$) and a weighted path-length ratio ($\bar{\Gamma}$)\cite{Blaney_JBO26_BroadlinewidthSources}:
\begin{equation}\label{equ:WmBLLcal}
	\acrshort{SaO2}(\acrshort{RoR},\acrshort{M},w) = \frac
	{-\bar{\epsilon}_{\acrshort{Hb},r}(\acrshort{M},w) + \bar{\epsilon}_{\acrshort{Hb},i}(\acrshort{M},w) \times \bar{\Gamma}(\acrshort{M},w) \times \acrshort{RoR}}
	{(\bar{\epsilon}_{\acrshort{HbO2},r}(\acrshort{M},w)-\bar{\epsilon}_{\acrshort{Hb},r}(\acrshort{M},w)) +  (\bar{\epsilon}_{\acrshort{Hb},i}(\acrshort{M},w)-\bar{\epsilon}_{\acrshort{HbO2},i}(\acrshort{M},w)) \times \bar{\Gamma}(\acrshort{M},w) \times \acrshort{RoR}} + 
	o
\end{equation}
where a weighted extinction coefficient is expressed as:
\begin{equation}\label{equ:Weps}
	\bar{\epsilon}(\acrshort{M},w)=
	\frac{\int I(\lambda,\acrshort{M},w) \epsilon(\lambda) d\lambda}
	{\int I(\lambda,\acrshort{M},w) d\lambda}
\end{equation}
where $I(\lambda,\acrshort{M})$ is the power-spectral density of the detected light which depends on \gls{M} due to spectral coloring if the light source is polychromatic\cite{Blaney_JBO26_BroadlinewidthSources}.
The weighted path-length ratio is expressed as:
\begin{equation}\label{equ:WGamma}
	\bar{\Gamma}(\acrshort{M},w)=\Gamma \times \gamma(\acrshort{M},w)
\end{equation}
where:
\begin{equation}
	\gamma(\acrshort{M},w) = \frac{\bar{\langle\ell\rangle}_i(\acrshort{M},w)/\bar{\langle\ell\rangle}_r(\acrshort{M},w)}{\Gamma_{mono}}
\end{equation}
and a weighted optical path-length ($\bar{\langle\ell\rangle}$) is:
\begin{equation}\label{equ:Well}
	\bar{\langle\ell\rangle}(\acrshort{M},w) = \frac{\int I(\lambda,\acrshort{M},w) \langle\ell\rangle(\lambda) d\lambda}
	{\int I(\lambda,\acrshort{M},w) d\lambda}
\end{equation}
where $\Gamma_{mono}$ and $\langle\ell\rangle$ are derived from the slab bulk tissue model presented in Ref.~\citenum{Blaney_JBO26_BroadlinewidthSources} and $\Gamma_{mono}$ is the modeled value for the nominal monochromatic wavelengths.
The purpose of the variable $\gamma$ is to capture the spectral coloring dependence of $\Gamma$ on \gls{M} using an analytical model, but $\Gamma$ (Equation~\ref{equ:WGamma}) is still directly fit for in case three.
This makes the dependence of $\bar{\Gamma}$ on \gls{M} model dependent, but not the amplitude of the fit parameter $\Gamma$.
The takeaway of case three (Equation~\ref{equ:WmBLLcal}) is that the detected spectrum depends on \acrfull{M}, making \gls{M} an independent variable in the calibration.
All three of these calibration cases contain two free parameters: 
Equation~\ref{equ:linCal} -- $a$ and $b$; 
Equations~\ref{equ:mBLLcal}~\&~\ref{equ:WmBLLcal} -- $\Gamma$ and $o$.
Importantly, the $\Gamma$ parameter has a physical interpretation\cite{Blaney_JBO24_CriticalAnalysis} while the interpretation of the empirical parameters---$a$, $b$, and $o$---is less clear.
\par

In Equation~\ref{equ:mBLLcal}, the nominal extinction coefficients used the values at \qtylist{660;910}{\nano\meter} for red and \gls{IR}, respectively.
In Equations~\ref{equ:WmBLLcal},~\ref{equ:Weps},~and~\ref{equ:Well}, the detected power-spectral density ($I$) was modeled as the product of the source emission power-spectral density ($P$), the epidermal transmission ($T_{epi}$), and the bulk-tissue transmittance Green's function ($G_{bulk}$)\cite{Blaney_JBO26_BroadlinewidthSources}:
\begin{equation}
	I(\lambda,\acrshort{M},w) = P(\lambda,w) \times T_{epi}(\lambda,\acrshort{M}) \times G_{bulk}(\lambda) \times \Lambda_{det}
\end{equation}
where $\Lambda_{det}$ is the area of the detector and $G_{bulk}$ was modeled using a \qty{15}{\milli\meter} thick slab as in Ref.~\citenum{Blaney_JBO26_BroadlinewidthSources}---the same model from which $\langle\ell\rangle$ values were calculated in Equation~\ref{equ:Well} (n.b., for computation a unit detector area was considered).
In this work, $P$ was modeled as a Gaussian broad-linewidth source:
\begin{equation}
	P(\lambda,w) = P_0 \times e^{-4\ln(2)\times\left(\lambda-\lambda_0\right)^2/w^2}
\end{equation}
where $w$ is the \gls{FWHM}, $\lambda_0$ is the nominal wavelength, and the expression is normalized so that $P(\lambda_0)=P_0$ where $P_0$ is the peak power-spectral density (n.b., for computation a unit peak power-spectral density was considered).
For the results in this work, we assumed a $w_i$ of \qty{50}{\nano\meter} for the \gls{IR} source (i.e., $\lambda_{0,i}=\qty{910}{\nano\meter}$) and a \gls{FWHM} of half that for the red source (i.e., $w_r$ of \qty{25}{\nano\meter} with $\lambda_{0,r}=\qty{660}{\nano\meter}$)\cite{Bierman_BJA24_MelaninBias,Blaney_JBO26_BroadlinewidthSources}.
At zero \gls{FWHM}---the monochromatic case---the weighted extinction coefficients reduce to their nominal values (i.e., same values as Equation~\ref{equ:mBLLcal}) and $\gamma = \num{1}$, so case three reduces to case two exactly when $w_r=w_i=\qty{0}{\nano\meter}$.
\par

To execute the calibration, case one (Equation~\ref{equ:linCal}) was fit by ordinary least squares. 
Cases two and three (Equations~\ref{equ:mBLLcal}~and~\ref{equ:WmBLLcal}) were fit by bounded nonlinear least squares---$\Gamma$ was bounded at \numrange{0.1}{10} and $o$ was bounded at \qtyrange{-100}{100}{\percent}.
A blood-draw's residual was defined as its recovered \gls{SaO2} using the calibrated model minus its measured \gls{SaO2}, and calibration accuracy was summarized as the root-mean-square residual, $A_{rms}$.
To test the calibration's dependence on skin tone, a linear regression considered the patient-mean residual versus patient pigment (i.e., \gls{M} or \gls{ITA}), weighting each patient by the inverse squared standard error of that patient's mean residual.
The quantity taken from this fit was its slope, in units of \gls{SaO2} percentage per unit pigment (e.g., \unit{\percent} per volume-fraction for \gls{M} and \unit{\percent} per \unit{\degree} for \gls{ITA}).
\par

This calibration procedure was carried out globally for the whole dataset and across \num{200} cross-validation folds.
For cross-validation, each fold drew a training set that satisfied the \gls{FDA} 2013 premarket-notification guidance for pulse oximeters\cite{FDACDRH_13_PulseOximeters}.
Following that guidance, a training set held \num{10} patients of both assigned sexes, \num{2} of them darkly pigmented---determined by their reported race (i.e., ``African American'' or ``Other/Multiethnic African American'') since the guidance does not define darkly pigmented---contributing at least \num{200} blood-draws inside the \qtyrange{70}{100}{\percent} \gls{SaO2} range.
Patients were drawn at random and redrawn until these conditions were met.
The \num{88} held-out patients (n.b., \num{98} total patients were considered for calibration) comprised the testing set which was used to determine the held-out $A_{rms}$ and held-out slope of \gls{SaO2} residual on pigment to test accuracy and skin-tone dependence, respectively (n.b., testing sets did consider blood-draws with $\acrshort{SaO2}<\qty{70}{\percent}$ while training sets did not).
This calibration step of the analysis is found in the \texttt{D\_calibrate\_openox.m} script in the GitHub repository\cite{Blaney_26_MelaninLinewidthcorrected}.
\par

\section*{Data availability}
The source dataset is the OpenOximetry Repository v1.1.1 on PhysioNet\cite{Fong_24_OpenOximetryRepository}, which is subject to the repository's data-use requirements.
No OpenOximetry data are redistributed with this manuscript.

\section*{Code availability}
The analysis code supporting this work is available at a public GitHub repository\cite{Blaney_26_MelaninLinewidthcorrected}.

\bibliography{refs260915}

\begin{thebibliography}{10}
\urlstyle{rm}
\expandafter\ifx\csname url\endcsname\relax
  \def\url#1{\texttt{#1}}\fi
\expandafter\ifx\csname urlprefix\endcsname\relax\def\urlprefix{URL }\fi
\expandafter\ifx\csname doiprefix\endcsname\relax\def\doiprefix{DOI: }\fi
\providecommand{\bibinfo}[2]{#2}
\providecommand{\eprint}[2][]{\url{#2}}

\bibitem{Aoyagi_JAnesth03_PulseOximetry}
\bibinfo{author}{Aoyagi, T.}
\newblock \bibinfo{journal}{\bibinfo{title}{Pulse oximetry: Its invention,
  theory, and future}}.
\newblock {\emph{\JournalTitle{Journal of Anesthesia}}}
  \textbf{\bibinfo{volume}{17}}, \bibinfo{pages}{259--266},
  \doiprefix\url{10.1007/s00540-003-0192-6} (\bibinfo{year}{2003}).

\bibitem{Charlton_Proc.IEEE22_WearablePhotoplethysmography}
\bibinfo{author}{Charlton, P.~H.} \emph{et~al.}
\newblock \bibinfo{journal}{\bibinfo{title}{Wearable {{Photoplethysmography}}
  for {{Cardiovascular Monitoring}}}}.
\newblock {\emph{\JournalTitle{Proceedings of the IEEE}}}
  \textbf{\bibinfo{volume}{110}}, \bibinfo{pages}{355--381},
  \doiprefix\url{10.1109/JPROC.2022.3149785} (\bibinfo{year}{2022}).

\bibitem{FDACDRH_13_PulseOximeters}
\bibinfo{author}{{FDA CDRH}}.
\newblock \bibinfo{title}{Pulse {{Oximeters}} - {{Premarket Notification
  Submissions}} [510(k)s]: {{Guidance}} for {{Industry}} and {{Food}} and
  {{Drug Administration Staff}}}.
\newblock
  \bibinfo{howpublished}{\url{https://www.fda.gov/regulatory-information/search-fda-guidance-documents/pulse-oximeters-premarket-notification-submissions-510ks-guidance-industry-and-food-and-drug}}
  (\bibinfo{year}{2013}).

\bibitem{SjodingMichaelW._NEJM20_RacialBias}
\bibinfo{author}{{Sjoding Michael W.}}, \bibinfo{author}{{Dickson Robert P.}},
  \bibinfo{author}{{Iwashyna Theodore J.}}, \bibinfo{author}{{Gay Steven E.}}
  \& \bibinfo{author}{{Valley Thomas S.}}
\newblock \bibinfo{journal}{\bibinfo{title}{Racial {{Bias}} in {{Pulse Oximetry
  Measurement}}}}.
\newblock {\emph{\JournalTitle{New England Journal of Medicine}}}
  \textbf{\bibinfo{volume}{383}}, \bibinfo{pages}{2477--2478},
  \doiprefix\url{10.1056/nejmc2029240} (\bibinfo{year}{2020}).

\bibitem{Shi_BMCMed.22_AccuracyPulse}
\bibinfo{author}{Shi, C.} \emph{et~al.}
\newblock \bibinfo{journal}{\bibinfo{title}{The accuracy of pulse oximetry in
  measuring oxygen saturation by levels of skin pigmentation: A systematic
  review and meta-analysis}}.
\newblock {\emph{\JournalTitle{BMC Medicine}}} \textbf{\bibinfo{volume}{20}},
  \bibinfo{pages}{267}, \doiprefix\url{10.1186/s12916-022-02452-8}
  (\bibinfo{year}{2022}).

\bibitem{Bickler_Ane.22_PulseOximeter}
\bibinfo{author}{Bickler, P.} \& \bibinfo{author}{Tremper, K.~K.}
\newblock \bibinfo{journal}{\bibinfo{title}{The {{Pulse Oximeter Is Amazing}},
  but {{Not Perfect}}}}.
\newblock {\emph{\JournalTitle{Anesthesiology}}}
  \textbf{\bibinfo{volume}{136}}, \bibinfo{pages}{670--671},
  \doiprefix\url{10.1097/ALN.0000000000004171} (\bibinfo{year}{2022}).

\bibitem{Cabanas_Sensors22_SkinPigmentation}
\bibinfo{author}{Cabanas, A.~M.}, \bibinfo{author}{{Fuentes-Guajardo}, M.},
  \bibinfo{author}{Latorre, K.}, \bibinfo{author}{Le{\'o}n, D.} \&
  \bibinfo{author}{{Mart{\'i}n-Escudero}, P.}
\newblock \bibinfo{journal}{\bibinfo{title}{Skin pigmentation influence on
  pulse oximetry accuracy: A systematic review and bibliometric analysis}}.
\newblock {\emph{\JournalTitle{Sensors}}} \textbf{\bibinfo{volume}{22}},
  \bibinfo{pages}{3402}, \doiprefix\url{10.3390/s22093402}
  (\bibinfo{year}{2022}).

\bibitem{Al-Halawani_Physiol.Meas.23_ReviewEffect}
\bibinfo{author}{{Al-Halawani}, R.}, \bibinfo{author}{Charlton, P.~H.},
  \bibinfo{author}{Qassem, M.} \& \bibinfo{author}{Kyriacou, P.~A.}
\newblock \bibinfo{journal}{\bibinfo{title}{A review of the effect of skin
  pigmentation on pulse oximeter accuracy}}.
\newblock {\emph{\JournalTitle{Physiological Measurement}}}
  \textbf{\bibinfo{volume}{44}}, \bibinfo{pages}{05TR01},
  \doiprefix\url{10.1088/1361-6579/acd51a} (\bibinfo{year}{2023}).

\bibitem{Martin_BJA24_EffectSkin}
\bibinfo{author}{Martin, D.} \emph{et~al.}
\newblock \bibinfo{journal}{\bibinfo{title}{Effect of skin tone on the accuracy
  of the estimation of arterial oxygen saturation by pulse oximetry: A
  systematic review}}.
\newblock {\emph{\JournalTitle{British Journal of Anaesthesia}}}
  \textbf{\bibinfo{volume}{132}}, \bibinfo{pages}{945--956},
  \doiprefix\url{10.1016/j.bja.2024.01.023} (\bibinfo{year}{2024}).

\bibitem{Chesley_RC22_RacialDisparities}
\bibinfo{author}{Chesley, C.~F.} \emph{et~al.}
\newblock \bibinfo{journal}{\bibinfo{title}{Racial {{Disparities}} in {{Occult
  Hypoxemia}} and {{Clinically Based Mitigation Strategies}} to {{Apply}} in
  {{Advance}} of {{Technological Advancements}}}}.
\newblock {\emph{\JournalTitle{Respiratory Care}}}
  \textbf{\bibinfo{volume}{67}}, \bibinfo{pages}{1499--1507},
  \doiprefix\url{10.4187/respcare.09769} (\bibinfo{year}{2022}).

\bibitem{Gudelunas_AA24_LowPerfusion}
\bibinfo{author}{Gudelunas, M.~K.} \emph{et~al.}
\newblock \bibinfo{journal}{\bibinfo{title}{Low {{Perfusion}} and {{Missed
  Diagnosis}} of {{Hypoxemia}} by {{Pulse Oximetry}} in {{Darkly Pigmented
  Skin}}: {{A Prospective Study}}}}.
\newblock {\emph{\JournalTitle{Anesthesia \& Analgesia}}}
  \textbf{\bibinfo{volume}{138}}, \bibinfo{pages}{552},
  \doiprefix\url{10.1213/ANE.0000000000006755} (\bibinfo{year}{2024}).

\bibitem{Hendrickson_CHESTCriticalCare26_EquiOxProspective}
\bibinfo{author}{Hendrickson, C.~M.} \emph{et~al.}
\newblock \bibinfo{journal}{\bibinfo{title}{{{EquiOx}}: {{A}} prospective study
  of pulse oximeter bias and skin pigmentation in critically-ill adults with
  stable oxygenation}}.
\newblock {\emph{\JournalTitle{CHEST Critical Care}}} \bibinfo{pages}{100305},
  \doiprefix\url{10.1016/j.chstcc.2026.100305} (\bibinfo{year}{2026}).

\bibitem{FDAARTDP_24_FDAExecutive}
\bibinfo{author}{{FDA ARTDP}}.
\newblock \bibinfo{title}{{{FDA Executive Summary}}: {{Performance Evaluation}}
  of {{Pulse Oximeters Taking}} into {{Consideration Skin Pigmentation}},
  {{Race}} and {{Ethnicity}}}.
\newblock \bibinfo{type}{Executive {{Summary}}}, \bibinfo{institution}{FDA}
  (\bibinfo{year}{2024}).
\newblock \bibinfo{note}{\url{https://www.fda.gov/media/175828/download}}.

\bibitem{Rea_BJA23_LightSource}
\bibinfo{author}{Rea, M.~S.} \& \bibinfo{author}{Bierman, A.}
\newblock \bibinfo{journal}{\bibinfo{title}{Light source spectra are the likely
  cause of systematic bias in pulse oximeter readings for individuals with
  darker skin pigmentation}}.
\newblock {\emph{\JournalTitle{British Journal of Anaesthesia}}}
  \textbf{\bibinfo{volume}{131}}, \bibinfo{pages}{e101--e103},
  \doiprefix\url{10.1016/j.bja.2023.04.018} (\bibinfo{year}{2023}).

\bibitem{Bierman_BJA24_MelaninBias}
\bibinfo{author}{Bierman, A.}, \bibinfo{author}{Benner, K.} \&
  \bibinfo{author}{Rea, M.~S.}
\newblock \bibinfo{journal}{\bibinfo{title}{Melanin bias in pulse oximetry
  explained by light source spectral bandwidth}}.
\newblock {\emph{\JournalTitle{British Journal of Anaesthesia}}}
  \textbf{\bibinfo{volume}{132}}, \bibinfo{pages}{957--963},
  \doiprefix\url{10.1016/j.bja.2024.01.037} (\bibinfo{year}{2024}).

\bibitem{Benner_JBO26_CauseEffect}
\bibinfo{author}{Benner, K.~J.}, \bibinfo{author}{Goel, N.~N.} \&
  \bibinfo{author}{Rea, M.~S.}
\newblock \bibinfo{journal}{\bibinfo{title}{Cause, effect, and remediation of
  melanin-associated bias in pulse oximetry}}.
\newblock {\emph{\JournalTitle{Journal of Biomedical Optics}}}
  \textbf{\bibinfo{volume}{31}}, \bibinfo{pages}{028002},
  \doiprefix\url{10.1117/1.JBO.31.2.028002} (\bibinfo{year}{2026}).

\bibitem{Benner_BritishJournalofAnaesthesia25_LightSource}
\bibinfo{author}{Benner, K.}, \bibinfo{author}{Bierman, A.} \&
  \bibinfo{author}{Rea, M.}
\newblock \bibinfo{journal}{\bibinfo{title}{Light source spectra and higher
  measurement uncertainty in pulse oximeter readings for individuals with
  darker skin pigmentation}}.
\newblock {\emph{\JournalTitle{British Journal of Anaesthesia}}}
  \textbf{\bibinfo{volume}{134}}, \bibinfo{pages}{856--858},
  \doiprefix\url{10.1016/j.bja.2024.11.034} (\bibinfo{year}{2025}).

\bibitem{Blaney_JBO26_BroadlinewidthSources}
\bibinfo{author}{Blaney, G.} \& \bibinfo{author}{Fantini, S.}
\newblock \bibinfo{journal}{\bibinfo{title}{Broad-linewidth sources result in a
  skin-tone bias in noninvasive optical measurement of oxygen saturation}}.
\newblock {\emph{\JournalTitle{Journal of Biomedical Optics}}}
  \textbf{\bibinfo{volume}{31}}, \bibinfo{pages}{070501},
  \doiprefix\url{10.1117/1.JBO.31.7.070501} (\bibinfo{year}{2026}).

\bibitem{Pologe_PLOSONE25_LaserbasedPulse}
\bibinfo{author}{Pologe, J.~A.}, \bibinfo{author}{You, N.~K.},
  \bibinfo{author}{Blumstein, M.}, \bibinfo{author}{Snyder, K.~L.} \&
  \bibinfo{author}{Hay, W.~W.}
\newblock \bibinfo{journal}{\bibinfo{title}{Laser-based pulse oximetry
  eliminates pigmentation effects on oxygen saturation measurements: {{A}}
  pilot study}}.
\newblock {\emph{\JournalTitle{PLOS ONE}}} \textbf{\bibinfo{volume}{20}},
  \bibinfo{pages}{e0333109}, \doiprefix\url{10.1371/journal.pone.0333109}
  (\bibinfo{year}{2025}).

\bibitem{Blaney_JBO24_DualratioApproach}
\bibinfo{author}{Blaney, G.}, \bibinfo{author}{Frias, J.},
  \bibinfo{author}{Tavakoli, F.}, \bibinfo{author}{Sassaroli, A.} \&
  \bibinfo{author}{Fantini, S.}
\newblock \bibinfo{journal}{\bibinfo{title}{Dual-ratio approach to pulse
  oximetry and the effect of skin tone}}.
\newblock {\emph{\JournalTitle{Journal of Biomedical Optics}}}
  \textbf{\bibinfo{volume}{29}}, \bibinfo{pages}{S33311},
  \doiprefix\url{10.1117/1.JBO.29.S3.S33311} (\bibinfo{year}{2024}).

\bibitem{Al-Halawani_JBO24_MonteCarlo}
\bibinfo{author}{{Al-Halawani}, R.}, \bibinfo{author}{Qassem, M.} \&
  \bibinfo{author}{Kyriacou, P.~A.}
\newblock \bibinfo{journal}{\bibinfo{title}{Monte {{Carlo}} simulation of the
  effect of melanin concentration on light--tissue interactions for
  transmittance pulse oximetry measurement}}.
\newblock {\emph{\JournalTitle{Journal of Biomedical Optics}}}
  \textbf{\bibinfo{volume}{29}}, \bibinfo{pages}{S33305},
  \doiprefix\url{10.1117/1.JBO.29.S3.S33305} (\bibinfo{year}{2024}).

\bibitem{Stuban_PPEE08_NoninvasiveCalibration}
\bibinfo{author}{Stub{\'a}n, N.} \& \bibinfo{author}{Masatsugu, N.}
\newblock \bibinfo{journal}{\bibinfo{title}{Non-invasive calibration method for
  pulse oximeters}}.
\newblock {\emph{\JournalTitle{Periodica Polytechnica Electrical Engineering
  (Archives)}}} \textbf{\bibinfo{volume}{52}}, \bibinfo{pages}{91--94},
  \doiprefix\url{10.3311/pp.ee.2008-1-2.11} (\bibinfo{year}{2008}).

\bibitem{Chan_RespiratoryMedicine13_PulseOximetry}
\bibinfo{author}{Chan, E.~D.}, \bibinfo{author}{Chan, M.~M.} \&
  \bibinfo{author}{Chan, M.~M.}
\newblock \bibinfo{journal}{\bibinfo{title}{Pulse oximetry: {{Understanding}}
  its basic principles facilitates appreciation of its limitations}}.
\newblock {\emph{\JournalTitle{Respiratory Medicine}}}
  \textbf{\bibinfo{volume}{107}}, \bibinfo{pages}{789--799},
  \doiprefix\url{10.1016/j.rmed.2013.02.004} (\bibinfo{year}{2013}).

\bibitem{Blaney_JBO24_CriticalAnalysis}
\bibinfo{author}{Blaney, G.}, \bibinfo{author}{Sassaroli, A.} \&
  \bibinfo{author}{Fantini, S.}
\newblock \bibinfo{journal}{\bibinfo{title}{Critical analysis of the
  relationship between arterial saturation and the ratio-of-ratios used in
  pulse oximetry}}.
\newblock {\emph{\JournalTitle{Journal of Biomedical Optics}}}
  \textbf{\bibinfo{volume}{29}}, \bibinfo{pages}{S33313},
  \doiprefix\url{10.1117/1.JBO.29.S3.S33313} (\bibinfo{year}{2024}).

\bibitem{Fong_24_OpenOximetryRepository}
\bibinfo{author}{Fong, N.}, \bibinfo{author}{Lipnick, M.},
  \bibinfo{author}{Feiner, J.} \& \bibinfo{author}{Law, T.}
\newblock \bibinfo{title}{{{OpenOximetry Repository}}},
  \doiprefix\url{10.13026/2g7z-t345} (\bibinfo{year}{2024}).

\bibitem{Wu_JBO23_SelfcalibratedPulse}
\bibinfo{author}{Wu, J.} \emph{et~al.}
\newblock \bibinfo{journal}{\bibinfo{title}{Self-calibrated pulse oximetry
  algorithm based on photon pathlength change and the application in human
  freedivers}}.
\newblock {\emph{\JournalTitle{Journal of Biomedical Optics}}}
  \textbf{\bibinfo{volume}{28}}, \bibinfo{pages}{115002},
  \doiprefix\url{10.1117/1.jbo.28.11.115002} (\bibinfo{year}{2023}).

\bibitem{Feiner_AA07_DarkSkin}
\bibinfo{author}{Feiner, J.~R.}, \bibinfo{author}{Severinghaus, J.~W.} \&
  \bibinfo{author}{Bickler, P.~E.}
\newblock \bibinfo{journal}{\bibinfo{title}{Dark {{Skin Decreases}} the
  {{Accuracy}} of {{Pulse Oximeters}} at {{Low Oxygen Saturation}}: {{The
  Effects}} of {{Oximeter Probe Type}} and {{Gender}}}}.
\newblock {\emph{\JournalTitle{Anesthesia \& Analgesia}}}
  \textbf{\bibinfo{volume}{105}}, \bibinfo{pages}{S18},
  \doiprefix\url{10.1213/01.ane.0000285988.35174.d9} (\bibinfo{year}{2007}).

\bibitem{Blaney_ProcSPIE25_PreliminaryInvestigation}
\bibinfo{author}{Blaney, G.}, \bibinfo{author}{Frias, J.},
  \bibinfo{author}{Tavakoli, F.}, \bibinfo{author}{Sassaroli, A.} \&
  \bibinfo{author}{Fantini, S.}
\newblock \bibinfo{title}{Preliminary investigation of partial optical
  path-lengths in pulse oximetry}.
\newblock In \emph{\bibinfo{booktitle}{Proc {{SPIE}}}}, vol.
  \bibinfo{volume}{13314}, \bibinfo{pages}{82--86},
  \doiprefix\url{10.1117/12.3042282} (\bibinfo{publisher}{SPIE},
  \bibinfo{year}{2025}).

\bibitem{Walton_JCMC10_MeasuringVenous}
\bibinfo{author}{Walton, Z.~D.}, \bibinfo{author}{Kyriacou, P.~A.},
  \bibinfo{author}{Silverman, D.~G.} \& \bibinfo{author}{Shelley, K.~H.}
\newblock \bibinfo{journal}{\bibinfo{title}{Measuring venous oxygenation using
  the photoplethysmograph waveform}}.
\newblock {\emph{\JournalTitle{Journal of Clinical Monitoring and Computing}}}
  \textbf{\bibinfo{volume}{24}}, \bibinfo{pages}{295--303},
  \doiprefix\url{10.1007/s10877-010-9248-y} (\bibinfo{year}{2010}).

\bibitem{Kainerstorfer_JBO16_OpticalOximetry}
\bibinfo{author}{Kainerstorfer, J.~M.}, \bibinfo{author}{Sassaroli, A.} \&
  \bibinfo{author}{Fantini, S.}
\newblock \bibinfo{journal}{\bibinfo{title}{Optical oximetry of
  volume-oscillating vascular compartments: Contributions from oscillatory
  blood flow}}.
\newblock {\emph{\JournalTitle{Journal of Biomedical Optics}}}
  \textbf{\bibinfo{volume}{21}}, \bibinfo{pages}{101408},
  \doiprefix\url{10.1117/1.jbo.21.10.101408} (\bibinfo{year}{2016}).

\bibitem{Matos_SciData24_BOLDBloodgas}
\bibinfo{author}{Matos, J.} \emph{et~al.}
\newblock \bibinfo{journal}{\bibinfo{title}{{{BOLD}}: {{Blood-gas}} and
  {{Oximetry Linked Dataset}}}}.
\newblock {\emph{\JournalTitle{Scientific Data}}}
  \textbf{\bibinfo{volume}{11}}, \bibinfo{pages}{535},
  \doiprefix\url{10.1038/s41597-024-03225-z} (\bibinfo{year}{2024}).

\bibitem{Blaney_26_MelaninLinewidthcorrected}
\bibinfo{author}{Blaney, G.} \& \bibinfo{author}{Frias, J.}
\newblock \bibinfo{title}{Melanin- and linewidth-corrected pulse-ox on the
  {{Open Oximetry}} dataset}.
\newblock
  \bibinfo{howpublished}{\url{https://github.com/gblane/melanin-linewidth-corrected-pulseox-on-openox}}
  (\bibinfo{year}{2026}).

\bibitem{Silva_JORS14_OpensourceToolbox}
\bibinfo{author}{Silva, I.} \& \bibinfo{author}{Moody, G.}
\newblock \bibinfo{journal}{\bibinfo{title}{An {{Open-source Toolbox}} for
  {{Analysing}} and {{Processing PhysioNet Databases}} in {{MATLAB}} and
  {{Octave}}}}.
\newblock {\emph{\JournalTitle{Journal of Open Research Software}}}
  \textbf{\bibinfo{volume}{2}}, \doiprefix\url{10.5334/jors.bi}
  (\bibinfo{year}{2014}).

\bibitem{Vasudevan_Comm.Med.24_MelanometryObjective}
\bibinfo{author}{Vasudevan, S.}, \bibinfo{author}{Vogt, W.~C.},
  \bibinfo{author}{Weininger, S.} \& \bibinfo{author}{Pfefer, T.~J.}
\newblock \bibinfo{journal}{\bibinfo{title}{Melanometry for objective
  evaluation of skin pigmentation in pulse oximetry studies}}.
\newblock {\emph{\JournalTitle{Communications Medicine}}}
  \textbf{\bibinfo{volume}{4}}, \bibinfo{pages}{1--19},
  \doiprefix\url{10.1038/s43856-024-00550-7} (\bibinfo{year}{2024}).

\bibitem{Kollias_ClinicsinDermatology95_PhysicalBasis}
\bibinfo{author}{Kollias, N.}
\newblock \bibinfo{journal}{\bibinfo{title}{The physical basis of skin color
  and its evaluation}}.
\newblock {\emph{\JournalTitle{Clinics in Dermatology}}}
  \textbf{\bibinfo{volume}{13}}, \bibinfo{pages}{361--367},
  \doiprefix\url{10.1016/0738-081x(95)00075-q} (\bibinfo{year}{1995}).

\bibitem{Yudovsky_J.Biophotonics11_RetrievingSkin}
\bibinfo{author}{Yudovsky, D.} \& \bibinfo{author}{Pilon, L.}
\newblock \bibinfo{journal}{\bibinfo{title}{Retrieving skin properties from in
  vivo spectral reflectance measurements}}.
\newblock {\emph{\JournalTitle{Journal of Biophotonics}}}
  \textbf{\bibinfo{volume}{4}}, \bibinfo{pages}{305--314},
  \doiprefix\url{10.1002/jbio.201000069} (\bibinfo{year}{2011}).

\bibitem{Jacques_PMB13_OpticalProperties}
\bibinfo{author}{Jacques, S.~L.}
\newblock \bibinfo{journal}{\bibinfo{title}{Optical properties of biological
  tissues: A review}}.
\newblock {\emph{\JournalTitle{Physics in Medicine and Biology}}}
  \textbf{\bibinfo{volume}{58}}, \bibinfo{pages}{R37--R61},
  \doiprefix\url{10.1088/0031-9155/58/11/r37} (\bibinfo{year}{2013}).

\bibitem{UCSFHypoxiaLab_25_ProtocolSkin}
\bibinfo{author}{{UCSF Hypoxia Lab}}.
\newblock \bibinfo{title}{Protocol for skin color assessment for pulse oximeter
  performance studies}.
\newblock
  \bibinfo{howpublished}{\url{https://openoximetry.org/study-protocols/}}
  (\bibinfo{year}{2025}).

\bibitem{KonicaMinolta_18_SpectrophotometerCM700d}
\bibinfo{author}{{Konica Minolta}}.
\newblock \bibinfo{title}{Spectrophotometer {{CM-700d}}/600d {{Instruction
  Manual}}}.
\newblock
  \bibinfo{howpublished}{\url{https://sensing.konicaminolta.us/wp-content/uploads/cm-700d_instruction_eng-54pn5p743t.pdf}}
  (\bibinfo{year}{2018}).

\bibitem{Blaney_JIOHS24_SpatialSensitivity}
\bibinfo{author}{Blaney, G.}, \bibinfo{author}{Sassaroli, A.} \&
  \bibinfo{author}{Fantini, S.}
\newblock \bibinfo{journal}{\bibinfo{title}{Spatial sensitivity to absorption
  changes for various near-infrared spectroscopy methods: {{A}} compendium
  review}}.
\newblock {\emph{\JournalTitle{Journal of Innovative Optical Health Sciences}}}
  \textbf{\bibinfo{volume}{17}}, \bibinfo{pages}{2430001},
  \doiprefix\url{10.1142/s1793545824300015} (\bibinfo{year}{2024}).

\bibitem{Garcia-Valenzuela_J.Phys.:Conf.Ser.11_AssessmentSaunderson}
\bibinfo{author}{{Garc{\'i}a-Valenzuela}, A.}, \bibinfo{author}{Cuppo, F.
  L.~S.} \& \bibinfo{author}{Olivares, J.~A.}
\newblock \bibinfo{journal}{\bibinfo{title}{An assessment of {{Saunderson}}
  corrections to the diffuse reflectance of paint films}}.
\newblock {\emph{\JournalTitle{Journal of Physics: Conference Series}}}
  \textbf{\bibinfo{volume}{274}}, \bibinfo{pages}{012125},
  \doiprefix\url{10.1088/1742-6596/274/1/012125} (\bibinfo{year}{2011}).

\bibitem{Berns_19_BillmeyerSaltzmans}
\bibinfo{author}{Berns, R.~S.}
\newblock \emph{\bibinfo{title}{Billmeyer and {{Saltzman}}'s {{Principles}} of
  {{Color Technology}}}} (\bibinfo{publisher}{John Wiley \& Sons},
  \bibinfo{address}{Hoboken, NJ USA}, \bibinfo{year}{2019}),
  \bibinfo{edition}{4} edn.

\end{thebibliography}

\section*{Acknowledgements}
G.B. is funded by the National Institutes of Health grant K99-HL181290.
The content is solely the authors’ responsibility and does not necessarily represent the official views of the awarding institutions.

\section*{Author contributions statement}
G.B. and V.K. conceived the study. 
G.B., J.F., and S.F. developed the theory. 
G.B. and R.D. curated the dataset. 
G.B. wrote the analysis code, preformed the analysis, drafted the manuscript, and acquired funding. 
S.F. and V.K. supervised the work. 
All authors reviewed the manuscript.

\section*{Additional information}
The authors declare no competing interests.

\end{document}